# Nanoscale engineering of optical strong coupling inside metals

Artyom Assadillayev[1], Ihar Faniayeu[2], Alexandre Dmitriev[2*], and Søren Raza[1*]

[1]Department of Physics, Technical University of Denmark, Fysikvej, DK-2800 Kongens Lyngby, Denmark.

[2]Department of Physics, University of Gothenburg, SE-412 96 Göteborg, Sweden

*Email: alexd@physics.gu.se; sraz@dtu.dk

**Abstract**

Optical polaritons appear when a material excitation strongly couples to the optical mode. Such strong coupling between molecular transitions and optical cavities results in far-reaching opportunities in modifying fundamental properties of chemical matter. More recently an exciting prospect of cavity-free polaritons has emerged by matter sustaining the optical mode with its geometry. Here we show how strong coupling of the interband transition and surface plasmons can be engineered in nickel at the nanoscale to realize cavity-free optical polaritons inside metals. Using electron energy-loss spectroscopy, we demonstrate that in thin films and nanoantennas the propagation and radiation losses result in a broadening of the plasmon linewidth and a transition from strong to weak coupling. Further, higher-order plasmon resonances couple to the interband transition, and the multipolar coupled states acquire the field profile of the plasmon. Our results provide a fundamental understanding of plasmon-interband coupling in metals and establish the base for the design of unforeseen photocatalytic and magneto-optical nanosystems.

**Introduction**

Light-matter polaritons arise due to coherent energy exchange between the confined electromagnetic field in resonators and the transitions in molecular systems. The associated remarkable modification of the energy levels offers exciting opportunities of tuning various fundamental properties of chemical matter by coupling to light modes. This so-called strong coupling regime holds a key potential in a broad range of fields, ranging from new device functionalities[1] to the modification of chemical reactions[2,3]. The fundamental requirement for strong coupling is in bringing the cavity-emitter interaction to such an extent that the coherent energy exchange between light and molecular transitions becomes greater than the individual decay rates. This can be achieved by exposing molecules to the electromagnetic field confined by a resonator. The most common optical resonators are photonic cavities, where the electromagnetic field is confined in microvolumes by metallic mirrors (e.g., Fabry-Pérot cavities or multilayer heterostructures), and polaritons emerge as collective excitations between the light modes and the ensemble of molecular emitters/absorbers[4,5]. Most recently a notion of cavity-free polaritons has emerged, where optical modes sustained by materials themselves self-couple to their own electronic or vibrational resonances[6]. Cavity-free polaritons have been observed in resonant excitonic (Lorentzian-type) materials such as perovskites[7], transition metal dichalcogenides[8–11] and dye molecules[12] with potential consequences for polaritonic chemistry, exciton transport, and modified material properties.

Special case represents metals in the visible-to-ultraviolet spectral range, where the electronic excitation under scrutiny is the interband transition. The optical response of plasmonic systems is well described by the simple Drude model for photon energies below the threshold of interband transitions[13]. For gold and silver, which are by far the most used in plasmonics, the onset of interband transitions leads to strong nonradiative decay. This may be traced to the spectrally broad nature of the interband transitions in these metals. In contrast, nickel features a spectrally localized interband transition, where the plasmon resonance and interband transition couple to form new hybrid plasmon-interband states[14–16]. These states display spectral features markedly different from the common Drude picture of metals and are reminiscent of the strong coupling regime of light-matter interaction characterized by a Rabi energy splitting[17,18]. The coupled regime is accessed when the plasmon and interband transition exchange energy at a rate faster than their intrinsic decay rates. The shape and size of the nickel material determines the plasmon resonance energy and linewidth and facilitates the engineering of the plasmon-interband self-coupling between the weak and the strong coupling regimes, resulting in the generation of metallic cavity-free optical polaritons.

We show that the plasmon-interband self-coupling in nickel thin films and colloidal nickel nanoantennas gives rise to spectral features reminiscent of both strong and weak coupling physics depending on the linewidth of the plasmon resonance. In particular, we show that propagating surface plasmon polaritons in nickel thin films couple to the interband transition for thick films $(t > 15 \text{ nm})$, while the coupling diminishes for thinner films as the short-range surface plasmon polariton (SPP) and its increasing propagation losses broadens the plasmon linewidth. Further, we demonstrate that spherical nanoparticles that in the quasistatic limit are dominated by a single localized surface plasmon resonance (LSPR) of electric dipole nature also experience an interband coupling, leading to the formation of two dipole-interband states. We find that this interband coupling affects also higher-order multipolar LSPRs, such as the quadrupole LSPR, but gradually transitions into the weak coupling regime due to radiation losses as the particle size increases. We experimentally demonstrate such nanoscale engineering of weak and strong coupling in nickel films and nanoparticles using electron energy-loss spectroscopy (EELS) performed in a scanning transmission electron microscope. EELS can measure the optical response in a broad spectral range with unprecedented spatial resolution[19,20], and has been exploited to study the near-field of a variety of plasmonic structures[21–23] as well as resonant dielectric nanostructures[24–27]. EELS has also been used to study strong coupling in different systems[28–32]. EELS is particularly helpful for nickel nanostructures, since they support LSPRs in the deep ultraviolet, which is difficult to access with light-based spectroscopy. Our experimental results are supported by a theoretical model for the resonance frequencies of thin films and spherical nanoparticles in the quasi-static limit as well as full-field EELS simulations. The former provides an intuitive physical picture of the plasmon-interband coupling in the strong coupling limit, while the latter accurately captures the experimental features in both the strong and weak coupling regimes. Our work provides a fundamental understanding of the nanoscale plasmon-interband engineering in metals as it can be extended to other plasmonic materials with spectrally localized interband transitions such as copper and aluminum. We foresee that such engineering might open new horizons in the nanoscale design of metallic photocatalytic and magneto-optical systems[33], where both plasmons and interband transitions play a decisive role in how such systems function with light.

**Results**

First, we consider a thin nickel film supporting the excitation of SPPs. Tabulated values for the complex permittivity of bulk nickel[34] $\varepsilon(\omega)$ are shown in Fig. 1a along with a Drude-Lorentz model capturing the absorption peak centered at 4.4 eV due to interband transitions. The simulated EELS signal from an electron penetrating a nickel film[35] of varying thickness $t$ is presented in Fig. 1b. For films with thicknesses $t > 15$ nm, we observe two distinct peaks in the simulated EELS signal. Their resonance energies are located to the low- and high-energy sides of the interband transition and do not vary significantly across different film thicknesses. It's tempting to assign these two EELS peaks to the excitation of the long-range and short-range SPPs commonly supported by thin metal films[36,37]. However, the high SPP momenta probed by the EELS simulation (and in the experiments) effectively uncouples the SPPs at each interface of the nickel film, and we should only observe a single EELS peak. The uncoupled SPPs for the nickel film thicknesses with $t > 15$ nm allows us to determine the SP frequency from the quasi-static relation

$$\varepsilon(\omega) + 1 = 0. \tag{1}$$

To highlight the role of interband transitions on the SP resonances, we consider a lossless Drude-Lorentz model for the permittivity $\varepsilon_{\text{DL}}(\omega) = 1 - \frac{\omega_\text{p}^2}{\omega^2} + \frac{G\omega_0^2}{\omega_0^2 - \omega^2}$, where $\omega_\text{p}$, $G$, and $\omega_0$ denote the plasma frequency, oscillator strength, and oscillator frequency, respectively. Inserting the Drude-Lorentz permittivity in Eq. (1), we find that two resonances are supported at the frequencies $\omega_\pm$

$$\omega_\pm^2 = \frac{1}{2}\left[\omega_\text{L}^2 + \omega_\text{D}^2 \pm \sqrt{(\omega_\text{L}^2 - \omega_\text{D}^2)^2 + 4\kappa^2\omega_\text{L}\omega_\text{D}}\right]. \tag{2}$$

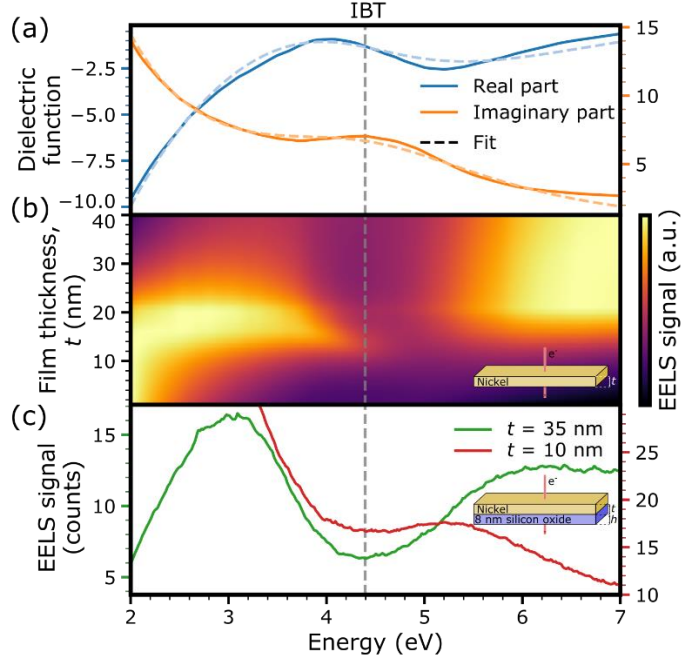

**Fig. 1 | Plasmon-interband cavity-free coupling in nickel thin films. (a)** Real and imaginary parts of the dielectric function of nickel along with a Drude-Lorentz fit. **(b)** Simulated EELS spectra of nickel thin films as a function of energy and film thickness $t$. **(c)** Experimental EELS spectra from nickel thin films with thicknesses $t=35$ nm (green line) and $t=10$ nm (red line) deposited on an 8 nm silicon dioxide membrane. The vertical dashed line indicates the interband transition energy at 4.4 eV.

Here, $\omega_D = \omega_p/\sqrt{2}$ is the well-known SP frequency for a Drude metal, while $\omega_L = \omega_0\sqrt{(G+2)/2}$ is the SP frequency for a pure Lorentz material. It is worth noting that Eq. (2) is of the same form as seen in coupled mechanical oscillator systems[38] mimicking strong-coupling phenomena in optical nanosystems. Indeed, $\kappa^2 = G\omega_L\omega_D/(G+2)$ is a coupling constant, which arises due to the interband transition, and determines the Rabi frequency splitting of the two SP resonances. These two coupled resonances are interpreted as hybrid plasmon-interband states due to the cavity-free coupling between the SPP and the interband transition in nickel.

As the film thickness decreases, the SPPs of the film hybridize and the EELS signal is dominated by the short-range SPP due to its strong confinement. The propagation losses of the short-range SPP increases with thinner films[39] and the Rabi splitting $\hbar\Omega$ is only observable if the following condition is met[40,41]

$$\hbar\Omega > \left[\frac{(\hbar\gamma_{SPP})^2}{2} + \frac{(\hbar\gamma_{IB})^2}{2}\right]^{\frac{1}{2}} \quad (3)$$

Here, $\hbar\gamma_{IB}$ and $\hbar\gamma_{SPP}$ denote the linewidths of the interband transition and SP resonance, respectively. Since both the interband and the plasmon are hosted by the same nickel nanostructure, their individual linewidths cannot be accessed. Nonetheless, the interband linewidth is constant and dictated by the nickel permittivity, while the SP linewidth can be controlled by the thickness of the nickel film. For films with thicknesses $t < 15$ nm, we observe that the increasing propagation losses (and, hence, increasing linewidth) of the short-range SPP decreases the interband-plasmon interaction and diminishes the Rabi splitting (Fig. 1b). In particular, we observe the formation of only a single EELS peak due to the short-range SPP as the systems exits the strong coupling regime. This peak shifts to lower energies with decreasing film thickness, which is consistent with the highly thickness dependent dispersion relation of the short-range SPP.

In Fig. 1(c), we present EELS measurements on nickel thin films deposited on a thin silicon dioxide transmission electron microscopy (TEM) membrane with two characteristic thicknesses experimentally visualizing the discussion above. For a film thickness of $t = 35$ nm, the interband transition and SP resonances are strongly coupled and two clear peaks can be observed in the EELS signal with a Rabi splitting energy of approximately 3 eV. As in the simulations (Fig. 1b), these two peaks are spectrally located at the low- and high-energy sides of the interband transition. For the thinner film ($t = 10$ nm), where SPP losses dominate and the plasmon-interband coupling is weak, we observe a single distinct peak due to the short-range SPP. This thickness-dependent tuning of the coupling is similar to that observed in *J*-aggregate-silver film coupled systems[42], but the crucial difference is that we observe it within the same cavity-free material.

The plasmon-interband coupling is not unique to the thin film case but applies for any nickel nanostructure. Figure 2 shows theoretical and experimental results for a small isolated nickel nanoparticle of radius $r = 21$ nm (see Methods for sample fabrication). The simulated EELS signal[43] from a nickel nanosphere in vacuum acquired at an impact parameter $b = 29$ nm shows two clear peaks (Fig. 2c), which are located to the low- and high-energy sides of the interband transition energy. We perform a multipole decomposition of the simulated EELS signal and find that the primary contribution to both EELS peaks is due to the dipole (i.e., multipolar order $l = 1$) LSPR. The surprising presence of two resonances due to a single multipolar order can be qualitatively understood by extending the quasi-static analysis from the thin film case to the multipolar Fröhlich resonance condition applicable to LSPRs

$$\varepsilon(\omega) + \frac{l+1}{l} = 0, \qquad (4)$$

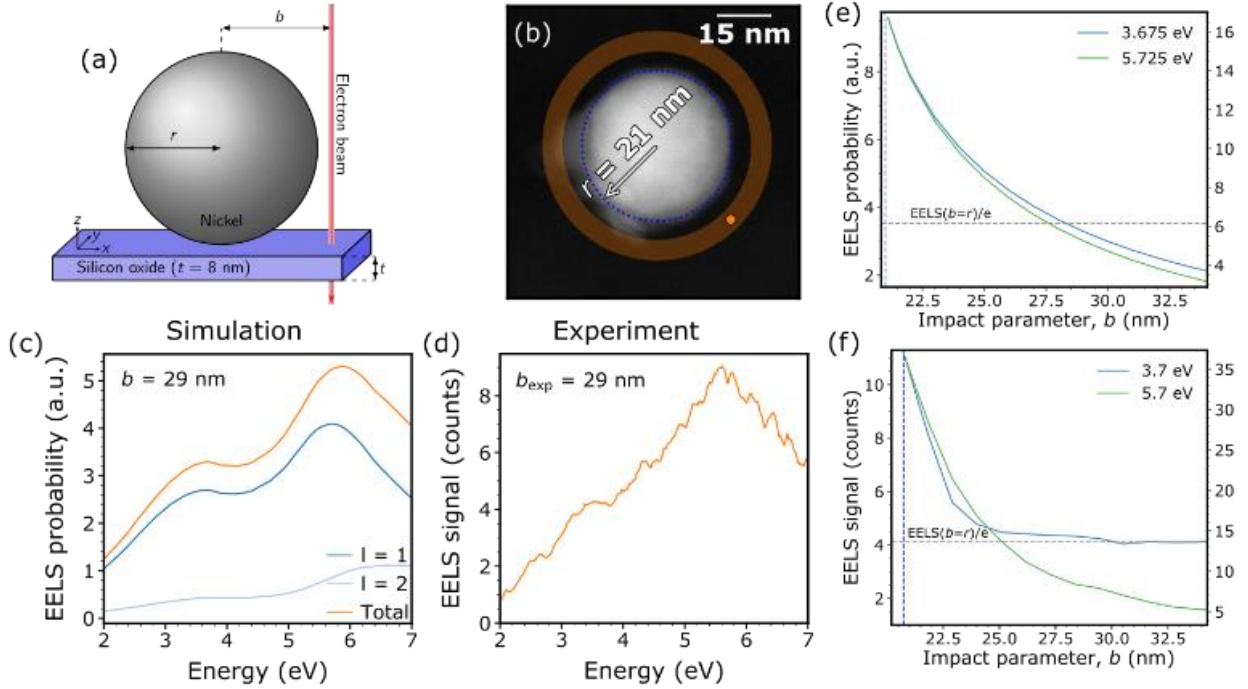

**Fig. 2 | Plasmon-interband polaritons in individual nickel nanoparticle. (a)** Schematic of a spherical nickel nanoparticle with radius *r* placed on a silicon oxide membrane probed by an electron beam with an impact parameter *b*. **(b)** STEM image of the nanoparticle, where the highlighted area represents the integration regions for the experimental EELS signal and the orange dot is the position of the electron beam in the simulation. **(c)** Simulated and **(d)** experimental EELS spectra of a nickel nanoparticle with *r* = 21 nm for the electron beam positions indicated in (b). **(e)** Simulated and **(f)** experimental EELS intensity spatial distribution at the two resonance energies of the nanoparticle. The grey dashed line represents the intensity value where the signal drops by a factor of *e* from its value at the particle edge.

where we have taken the background permittivity to be unity. Inserting the lossless Drude-Lorentz permittivity into Eq. (4), we arrive at a multipolar-dependent coupling of the LSPRs to the interband transition, given by

$$\omega_{\pm,l}^2 = \frac{1}{2}\left[\omega_{L,l}^2 + \omega_{D,l}^2 \pm \sqrt{(\omega_{L,l}^2 - \omega_{D,l}^2)^2 + 4\kappa_l^2 \omega_{L,l}\omega_{D,l}}\right]. \quad (5)$$

In Eq. (5), $\omega_{D,l} = \omega_p/L$ and $\omega_{L,l} = \sqrt{G+L^2}\,\omega_0/L$ denote the multipolar LSPR frequencies for a pure Drude and pure Lorentz material, respectively. Here, we have introduced the multipolar constant $L^2 = (2l+1)/l$, which for the dipole mode reduces to the familiar $L = \sqrt{3}$. The coupling constant also takes on a multipolar dependence and is given by $\kappa_l^2 = G\omega_{L,l}\omega_{D,l}/(G+L^2)$. Equation (5) reveals that each multipolar plasmon experiences an interband coupling and that the coupling strength is *l*-dependent.

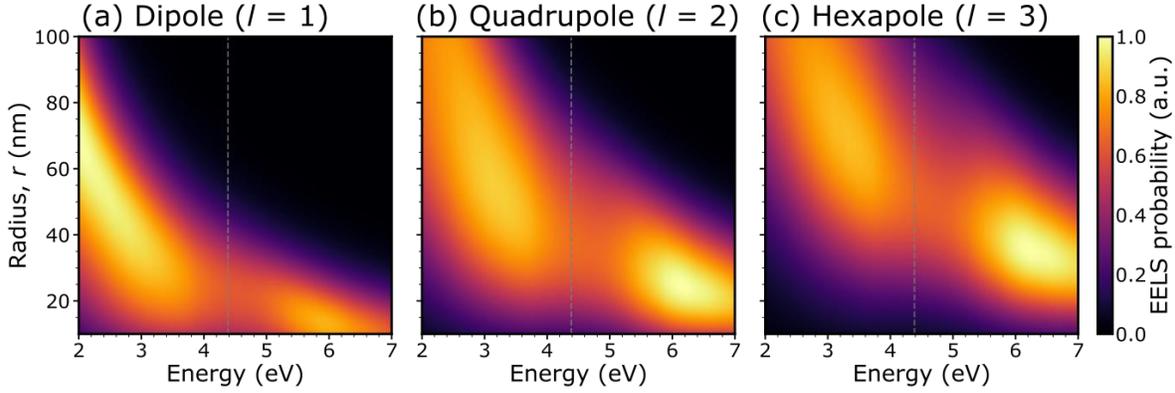

**Fig. 3 | Multipolar plasmon-interband cavity-free coupling.** EELS simulation of spherical nickel nanoparticles with varying radius $r$ decomposed into the multipolar orders **(a)** $l = 1$ (dipole), **(b)** $l = 2$ (quadrupole), and **(c)** $l = 3$ (hexapole). The impact parameter is $b = 3r$. The vertical dashed line indicates the interband transition energy at 4.4 eV.

The qualitative analysis provided by Eqs. (4-5) explains the presence of two resonances in the simulated EELS peaks, even though the particle only hosts the dipole mode (Fig. 2c). The experimental EELS spectrum (Fig. 2d) for the same impact parameter also shows two peaks, albeit with the high-energy peak (5.7 eV) more prominent than the low-energy peak (3.7 eV). We attribute this difference to the effect of the thin (8 nm) silicon dioxide substrate (see Supplementary Figure S1). The simulated and experimental beam-position-dependent EELS intensity profiles at the energies of the low- and high-energy peaks are shown in Fig. 4e-f, respectively. Here, we observe that both resonances are tightly confined to the surface of the particle with a $1/e$ drop in EELS intensity within a distance of approximately 5 nm. The experimental EELS profile for the low-energy peak (3.7 eV) reaches a baseline EELS value, which is effectively the noise level of our deconvoluted EELS data (see Methods), and therefore deviates from the simulated profile. Nonetheless, the position-dependent EELS profiles show that both coupled plasmon-interband states acquire a dipole-like field profile with a similar strong confinement to the particle surface. This is rather counterintuitive as interband transitions are typically considered a bulk property of the material. Yet, here we observe that the cavity-free coupling endows the dipolar profile of the LSPR to both of the plasmon-interband polaritons.

Higher-order multipoles ($l > 1$) become prominent as the particle size increases. We examine the multipole-dependent plasmon-interband coupling by calculating the EELS signal of a nickel nanosphere with increasing radii and decompose the EELS signal into separate multipole contributions (Fig. 3). The EELS signal is calculated using an analytical formula, which is valid for a spherical particle in vacuum and external impact parameters ($b > r$)[43]. The multipole-decomposed EELS signal shows that each multipole couples to the interband transition, which is in agreement with the analysis of Eqs. (4-5). In particular, increasing multipole order require larger particle sizes to match the LSPR energy to the interband transition (at 4.4 eV). Interestingly, we observe that increasing multipole order also shows weaker interband coupling, i.e., the Rabi splitting becomes less pronounced. This effect can, as for the thin film case (Fig. 1), be attributed to an increase in the plasmon linewidth. The larger size of the particles needed to sustain higher-order multipoles opens up a new loss channel, namely, radiation losses, which increases the plasmon linewidth and subsequently decreases the Rabi splitting, see Eq. (3). Consequently, the particle size strongly determines the nature of the optical response, where small particles close to the quasi-static limit are described by a strong-coupling-type response producing upper and lower plasmon-interband polaritons, while larger particles are closer to the weak coupling regime. In the latter, the interband transition acts primarily as a non-radiative loss channel for the LSPRs as in the case of noble metals.

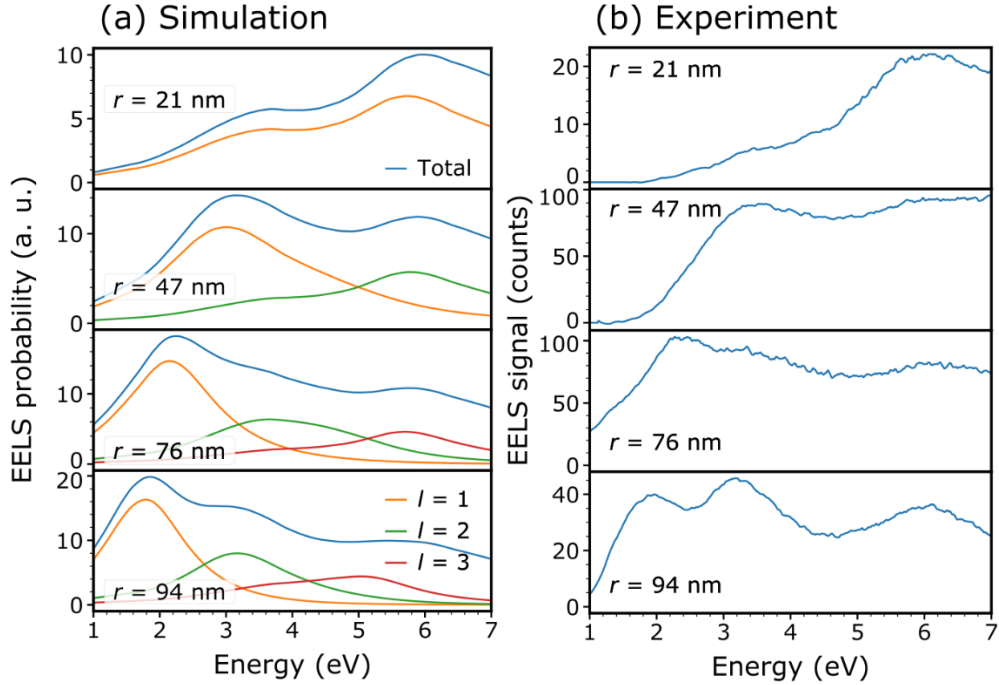

**Fig. 4 | Strong-to-weak plasmon-interband coupling. (a)** EELS simulations and **(b)** measurements for nickel nanoparticles with increasing radii for the impact parameters 24, 54, 88 and 109 nm, respectively. The simulations are decomposed into different multipole contributions for each particle size.

We examine this transition from strong to weak plasmon-interband coupling by simulating and measuring the EELS response of individual nickel nanoparticles with increasing size (Fig. 4). The simulated EELS spectra are decomposed into their main multipole contributions (Fig. 4a). We observe that the EELS signal of the smallest particle is primarily composed of the coupled dipole-interband states (as in Fig. 2). As the particle size increases $(r = 47 \text{ nm})$, the dipole LSPR shifts to energies lower than the interband transition, producing a single uncoupled peak. At the same time, we observe that the quadrupole LSPR $(l = 2)$ has redshifted in energy to induce an interband coupling. Such higher-order plasmon coupling has also been observed in plasmon-exciton systems.[44] With further increase in the particle size ($r = 76$ nm and $r = 94$ nm), the quadrupole LSPR redshifts to energies lower than the interband transitions and produces, as for the dipole case, a single uncoupled peak in the EELS signal. In these large particles, the hexapole LSPR ($l = 3$) also contributes to the EELS spectrum, but couples weaker to the interband transition. The weaker coupling is evidenced by both the smaller energy splitting as well as the asymmetric amplitudes of the two hexapole-interband states. These results are consistent with the discussion of Fig. 3 and indicate that each EELS peak in larger nickel nanoparticles can be assigned to individual (uncoupled) multipolar orders. In Fig. 4b, we show the measured EELS spectra from nickel nanoparticles of the same size and obtained from a similar impact parameter as in the simulations. The measured spectra are in overall good agreement with the simulations, demonstrating that the EELS response of nickel nanoparticles transitions from the plasmon-interband coupling to a largely uncoupled response as the particle size increases. We attribute this transition to the increased radiation losses of larger particles, which broadens the multipolar plasmon linewidths.

**Conclusion**

Cavity-free polaritons offer an exciting platform for leveraging the opportunities of strong coupling physics. Here, we have shown experimentally and theoretically that cavity-free polaritons emerge due to the coupling of the spectrally localized interband transition and the surface plasmons in nickel thin films and nanoantennas. We find that the plasmon linewidth dictates the nature of the plasmon-interband coupling and that the coupling can be engineered through the size of the nickel nanostructure. We observe a transition from strong coupling for thick films and small nanoparticles to weak coupling for thin films and larger nanoparticles. For both geometries, the strong-to-weak transition is initiated by an increase in the plasmon linewidth. Our work provides a fundamental understanding of cavity-free plasmon-interband polaritons in nickel, which is particularly relevant for nanoscale magneto-optics and magnetophotonics. The study can also be extended to other metals with spectrally localized interband transitions, such as aluminum and copper, and in the general design of metallic polaritons. We foresee, for example, that the metallic photocatalytic systems relying on interband transitions could gain exciting new functionalities if the dipolar or multipolar spatial profile of the emerging plasmon-interband polaritons considered in the design.

## Methods

### Simulations

The EELS simulations of nickel thin films presented in Fig. 1 are conducted using the analytical formalism described in Ref. [35] with a cut-off wavevector of $k_c = 6.03$ nm$^{-1}$ corresponding to a collection angle of $\theta_c = 3$ mrad appropriate for our experimental setup. The EELS simulations of spherical nickel nanoparticles presented in Figs. 2-4 are performed using the multipolar formalism derived in Ref. [43]. Tabulated values for the permittivity of nickel[34] are used in the calculations and in the Drude-Lorentz model given by

$$\varepsilon(\omega) = \varepsilon_\infty - \frac{\omega_p^2}{\omega(\omega + i\gamma)} + \frac{G\omega_0^2}{\omega_0^2 - \omega^2 - i\omega\gamma_{IB}}$$

The best fit to the experimental permittivity of nickel in the energy range 2-7 eV is given by the values: $\varepsilon_\infty = 2.61$, $\omega_p = 10.36$ eV, $\gamma = 1.61$ eV, $G = 3.45$, $\omega_0 = 4.72$ eV, and $\gamma_{IB} = 3.44$ eV.

### Sample fabrication

We use commercial nickel nanospheres of 99.7% purity in nanopowder form with a radius range of 40-180 nm. We add 1 g of the nanopowder to acetone and ultrasonicate for 10 min to ensure separation of the nanoparticles. The colloidal nickel nanoparticles in acetone are finally dropcasted onto an 8 nm thick silicon dioxide membrane for EELS measurements. We use the commercial evaporator Lesker PVD 225 for deposition of thin nickel films on top of with 8 nm thick silicon dioxide membranes with thicknesses of 1, 3, 5, 10, and 35 nm, respectively. The deposition process was carried out at the pressure of 5·10$^{-7}$ Torr with a deposition rate of 0.3 nm/s.

### Electron energy loss spectroscopy

The EELS measurements are performed in a monochromated and aberration-corrected FEI Titan operated in STEM mode at an acceleration voltage of 300 kV, providing a probe size of 0.5 nm and an energy resolution of 0.08 eV (as measured by the full-width-at-half-maximum of the zero-loss peak). We perform Richardson-Lucy deconvolution to remove the zero-loss peak using an EELS spectrum recorded in vacuum as the input for the point-spread function. Due to a small asymmetry in the zero-loss peak, the deconvolution algorithm produced an artificial EELS peak in the energy range below 1 eV. However, the artificial peak did not overlap with any of the observed resonances and could be safely removed using a first-order logarithmic polynomial. The STEM image analysis and EELS data analysis follows the same procedure as in our previous works.[24,25,45]

**Supporting Information**

Supporting Information contains additional EELS simulations.

**Acknowledgements**

A. A. and S. R. acknowledges support by the Independent Research Funding Denmark (7026-00117B). A. D. and I. F. acknowledge the Swedish Research Council VR (2017-04828) and Swedish Research Council for Sustainable Development FORMAS (2021-01390).

**Author contributions**

A. A. conducted the EELS measurements, data analysis, and EELS simulations. I. F. prepared the samples. A. D. and S. R. supervised the work. All authors contributed to the data interpretation and writing of the manuscript.

**Conflict of interest**

The authors declare no competing financial interests.

Supporting Information for

# Nanoscale engineering of optical strong coupling inside metals

Artyom Assadillayev[1], Ihar Faniayeu[2], Alexandre Dmitriev[2*], and Søren Raza[1*]

[1]Department of Physics, Technical University of Denmark, Fysikvej, DK-2800 Kongens Lyngby, Denmark.

[2]Department of Physics, University of Gothenburg, SE-412 96 Göteborg, Sweden

* Email: alexd@physics.gu.se; sraz@dtu.dk

The supporting information contains Supplementary Figure S1 in support of the main body of the text.

**Supporting Figures**

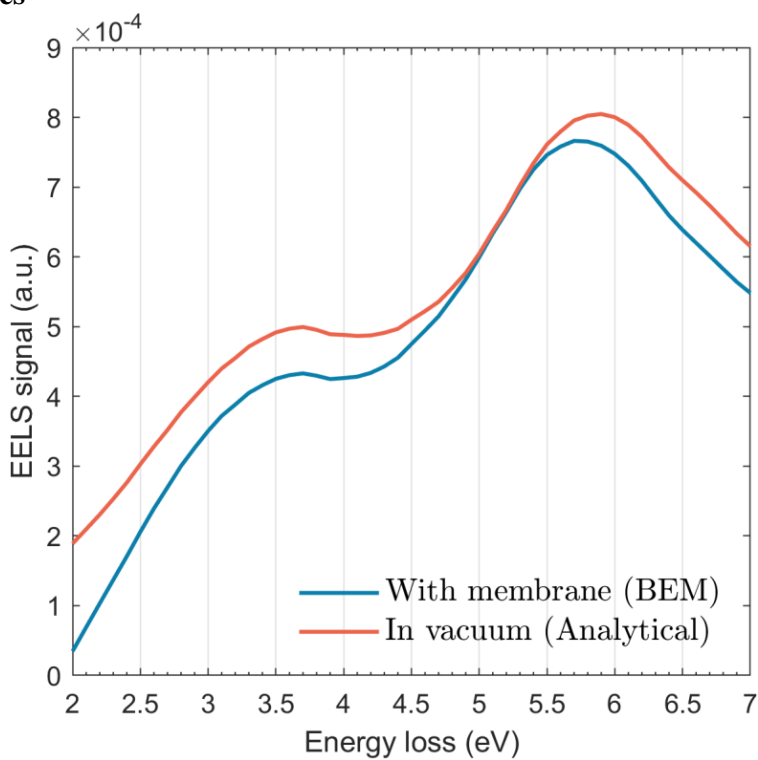

**Supplementary Fig. S1 | EELS simulation with and without membrane.** Simulated EELS signal from a nickel nanosphere with radius $r = 21$ nm in vacuum (red) and on a 8-nm-thick silicon dioxide membrane (blue) with an impact parameter of $b = 29$ nm in both cases. A gap of 1 nm between the particle and membrane is used to avoid a singular point in the simulation. The calculation with the membrane is performed using the boundary element method implemented in the MNPBEM toolbox for MATLAB.[1] The calculation in vacuum is performed using an analytical solution based on Mie theory for electron beam excitation.[2] Frequency-dependent refractive indices for silicon dioxide[3] and nickel[4] are employed. The simulations show that the membrane decreases the intensity of the lower plasmon-interband polariton, which is in agreement with our experimental observation in Fig. 2.

**Supporting references**